\documentclass[journal=nalefd, manuscript=letter, layout=twocolumn]{achemso}

\usepackage[version=3]{mhchem} 
\usepackage{multicol}
\usepackage{amsmath}             
\usepackage{amssymb}             
\usepackage{graphicx}            
\usepackage{array,booktabs}      
\usepackage{caption, float}      

\author{Zhongwei Dai}
\email{zbr5@wildcats.unh.edu}
\affiliation{Department of Physics and Materials Science Program, University of New Hampshire, Durham, New Hampshire 03824, USA}
\author{Wencan Jin}
\altaffiliation
{Current address: Department of Physics, University of Michigan, Ann Arbor, Michigan 48109, USA}
\affiliation{Columbia University, New York, New York 10027, USA}

\author{Jie-Xiang Yu}
\affiliation{Department of Physics and Materials Science Program, University of New Hampshire, Durham, New Hampshire 03824, USA}

\author{Maxwell Grady}
\affiliation{Department of Physics and Materials Science Program, University of New Hampshire, Durham, New Hampshire 03824, USA}

\author{Jerzy T. Sadowski}
\affiliation{Center for Functional Nanomaterials, Brookhaven National Laboratory, Upton, New York 11973, USA}

\author{Young Duck Kim}
\altaffiliation
{Current address: Department of Physics and Center for Humanities and Sciences, Kyung Hee University, Seoul 02447, Republic of Korea}
\author{James Hone}
\affiliation{Columbia University, New York, New York 10027, USA}

\author{Jiadong Zang}
\affiliation{Department of Physics and Materials Science Program, University of New Hampshire, Durham, New Hampshire 03824, USA}

\author{Richard M. Osgood, Jr.}
\affiliation{Columbia University, New York, New York 10027, USA}

\author{Karsten Pohl}
\affiliation{Department of Physics and Materials Science Program, University of New Hampshire, Durham, New Hampshire 03824, USA}

\title{
Surface buckling of phosphorene materials: determination, origin and influence on electronic structure\\}

\abbreviations{(BP), black phosphorus; (FLP) few-layer phosphorene; (LEEM), low energy electron microscopy; ($\mu$LEED) selected area micro-spot low energy electron diffraction; (PEEM) photo-emission electron microscopy}
\begin{document}
\begin{abstract}
The surface structure of phosphorene crystals materials is determined using surface sensitive dynamical micro-spot low energy electron diffraction ($\mu$LEED) analysis using a high spatial resolution low energy electron microscopy (LEEM) system. Samples of (\textit{i}) crystalline cleaved black phosphorus (BP) at 300 K and (\textit{ii}) exfoliated few-layer phosphorene (FLP) of about 10 nm thicknes, which were annealed at 573 K in vacuum were studied. In both samples, a significant surface buckling of 0.22 \AA\ and 0.30 \AA, respectively, is measured, which is one order of magnitude larger than previously reported. Using first principle calculations, the presence of surface vacancies is attributed not only to the surface buckling in BP and FLP, but also the previously reported intrinsic hole doping of phosphorene materials.
\end{abstract}

Black phosphorus (BP), which in its monolayer version is denoted as phosphorene, has recently had a rebirth as a new member of the family of the vigorously studied two-dimensional (2D) materials. It has attracted much attention due to its intriguing potential applications for modern electronics \cite{bp1,bp2,bp3,bp5} and photonics \cite{photo1,photo2}. For example, BP exhibits an intrinsic layer-dependent bandgap ranging from 0.3 eV (bulk) to 2 eV (monolayer) \cite{bandgap}, which bridges the energy gap between graphene and transition metal dichalcogenides (TMDs) \cite{bpreview}. This strong layer-dependent band gap presents a potential for an interesting set of integrated devices on a single supporting platform. Despite this surge of research in the applications of phospherene materials, much remains to be learned of its basic physical properties both from a device and a fundamental physics perspective. For example, its basic surface lattice structure is not well known, but is crucial both for predicting electronic properties and exploring phosphorene based devices.

\begin{figure}
	\begin{center}
\includegraphics[width=0.5\textwidth]{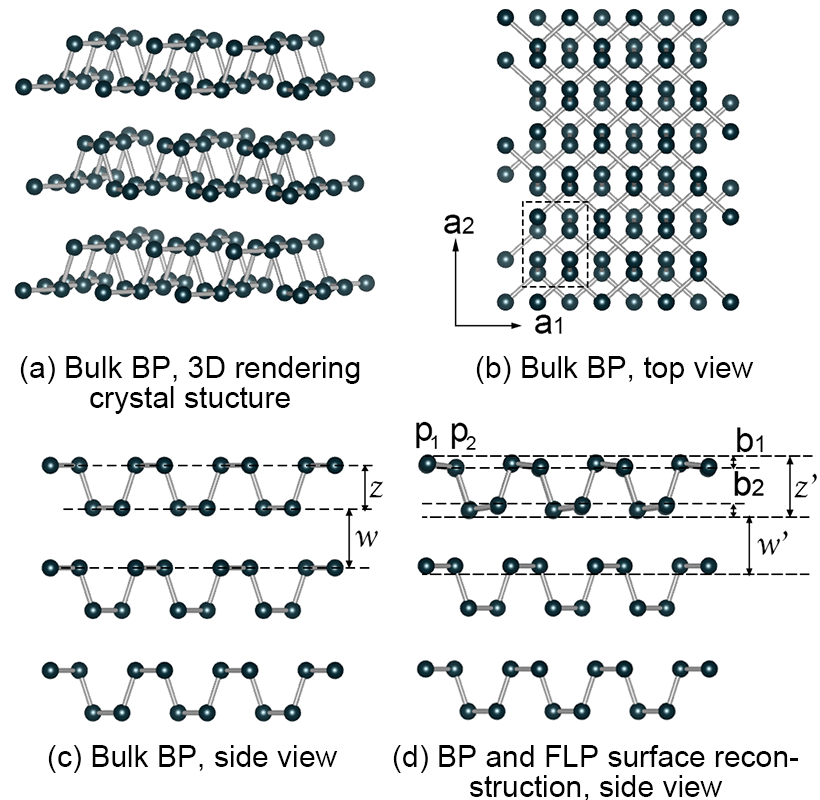}
\caption{(a), (b), (c): BP bulk crystal structure. (d): sideview of BP and FLP relaxed surface structure. Dotted square in (b) indicates the unit cell of BP, containing 8 P atoms.}
\label{fig:bulk}
	\end{center}
\end{figure}

In fact to date, there is no consensus on the atomic structure of surface region of BP. The crystal structure of BP, as shown in Fig. \ref{fig:bulk}, has a puckered honeycomb structure similar to that of graphene \cite{morita}. Two previous STM studies of phosphorene \cite{stm1,stm2} have revealed some aspects of the BP surface topography and observed an apparent height difference between the two symmetrically equivalent atoms P$_1$ and P$_2$, as shown in Fig. \ref{fig:bulk} (d). While these STM measurements were not able to quantify the geometrical height difference between P$_1$ and P$_2$, denoted as surface buckling, these studies did propose very small surface buckling values, 0.02 \AA\ \cite{stm1} and  0.06 \AA\ \cite{stm2}, based on first-principles calculations. In order to experimentally resolve the surface atomic structure of BP, there are two main challenges for the characterization technique: it has to be (\textit{i}) non-destructive and sensitive to the 3D atomic structure in the first few layers, and (\textit{ii})  able to restrict the sampling area to a few $\mu$m because many 2D materials including phosphorene are commonly prepared as very limited area exfoliated flakes. Selected area micro-spot low energy electron diffraction ($\mu$LEED) in a low energy electron microscope (LEEM), combined with dynamical intensity versus incoming electron energy (LEED-\textit{IV}) calculations, is one of the very few practical techniques to determine the 3D surface structure and composition of 2D materials with atomic resolution \cite{daileed, snse, ruleed, sunleed, sunleed2}.

In this letter, we present the first detailed experimental atomic surface structure determination of  BP. We use LEEM and dynamical $\mu$LEED-\textit{IV} analysis to examine in-situ-cleaved bulk BP surface and mechanically-exfoliated few-layer phosphorene (FLP) flakes of about 10 nm thickness. We measure the surface buckling for the two studied systems to be 0.22 \AA\ and 0.30 \AA, respectively, which are one order of magnitude larger than two previously reported theoretical values. Finally we use first-principles calculations to identify that the presence of surface vacancies is very likely the origin of not only the surface buckling in BP, but also the intrinsic hole doping of phosphorene that was reported previously \cite{bp3, ptype-1}.

\begin{figure}
\centering     
\includegraphics[width=80mm]{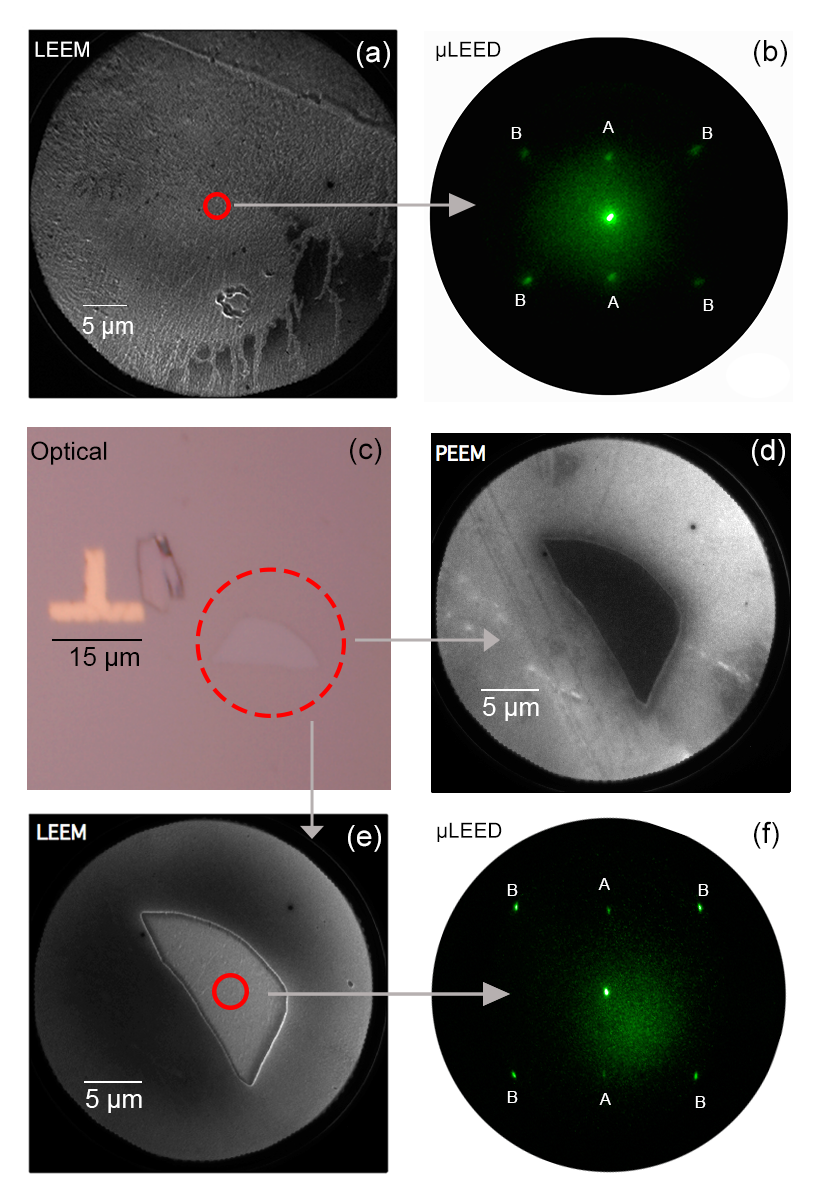}
\caption{(a) LEEM image taken at 30 eV electron energy and (b) $\mu$LEED diffraction pattern of freshly cleaved BP crystal surface. (c) Optical, (d) PEEM, (e) LEEM and (f) $\mu$LEED image of mechanically exfoliated flake of FLP, of about 10 nm thickness.}\label{fig:leed-leem}
\end{figure}

Our experiments were carried out on the Elmitec AC-LEEM system at the Center for Functional Nanomaterials in Brookhaven National Laboratory. The spatial resolution in LEEM mode is better than 3 nm and the electron beam spot size is 2 $\mu$m in diameter in the $\mu$LEED mode. Single-crystal bulk BP was cleaved in ultrahigh vacuum at room temperature. Figure \ref{fig:leed-leem} (a) shows the real-space bright field LEEM image of a freshly cleaved BP surface. $\mu$LEED data were acquired at the region denoted by the red 2 $\mu$m circle using a normal incident electron beam. Figure \ref{fig:leed-leem} (b) shows the well defined LEED pattern at 35 eV electron energy, indicating a very well-ordered surface. To prepare for the FLP samples, black phosphorous flakes were mechanically exfoliated onto n-doped Si chip with native oxide, using a previously described method \cite{paper3, mos2prb}. The substrate was pre-patterned with gold marks, which allowed us to locate and characterize the flakes of interest using optical microscope; see Fig. \ref{fig:leed-leem} (c). This procedure was performed in a Ne atmosphere. Subsequently, the sample was encapsulated and transferred to the LEEM chamber. The total exposure time of the exfoliated sample to air was less than 5 minutes. Even with such a short exposure time, significant surface oxidization and contamination was observed using photo-emission electron microscopy (PEEM). In order to remove the contamination, we annealed the sample at 300$^{\circ}$C in ultrahigh vacuum for 2 hours. The correct annealing temperature is crucial for successful contamination removal; higher temperatures will lead to accelerated sublimation. As shown in the PEEM and  LEEM images, Fig. \ref{fig:leed-leem} (d)-(e), the surface was pristine and uniform after successful annealing. $\mu$LEED data were acquired at the region denoted by the red circle using normal incident electrons. Figure \ref{fig:leed-leem} (f) shows the sharp LEED pattern at 25 eV electron energy, indicating a very well-ordered layered structure. To fully investigate the surface atomic structure we collected $\mu$LEED-\textit{IV} spectra for 7 recorded diffraction spots with an energy range of 25 to 135 eV for both sample varieties.  The intensities of symmetrically equivalent beams were averaged to minimize intensity anisotropy of the diffraction beam due to small sample titling ($<$$~$0.1$^{\circ}$). Specifically, as shown in Fig. \ref{fig:leed-leem} (b) and (f), intensities of spots A were averaged to assign the (01) diffraction beam and beam intensities of spots B were averaged to assign the (11) diffraction beam. The background intensity was then subtracted from the diffraction beam intensity.

Dynamical LEED-\textit{IV} analysis were carried out to extract the surface atomic structural information for bulk BP and FLP from the corresponding $\mu$LEED-\textit{IV} curves. In a dynamical LEED-\textit{IV} analysis, \textit{IV} curves are calculated for a trial structure and compared with experimental curves. A $\chi^2$-based $R_2$ reliability factor is used to quantify the difference between calculated and experimental \textit{IV} curves \cite{adam}. The surface structural parameters are then adjusted to search for the optimized surface structure that minimizes the $R_2$ factor. For electrons with an energy range of 25-135 eV, the mean free path is about 5 to 10 \AA. Use of this energy range means that our $\mu$LEED-\textit{IV} curves are most sensitive to the surface structure of the top few atomic layers of our samples, i.\ e.\, the buckling of the top atomic layer \textit{$b_1$}, the thickness of the first phosphorene layer \textit{$z$}', the buckling of the bottom atomic layer \textit{$b_2$} and the Van der Waals gap between the top and second phosphorene layer  \textit{w}', as demonstrated in Fig. \ref{fig:bulk} (d).

\begin{figure*}[]
\centering     
\includegraphics[width=160mm]{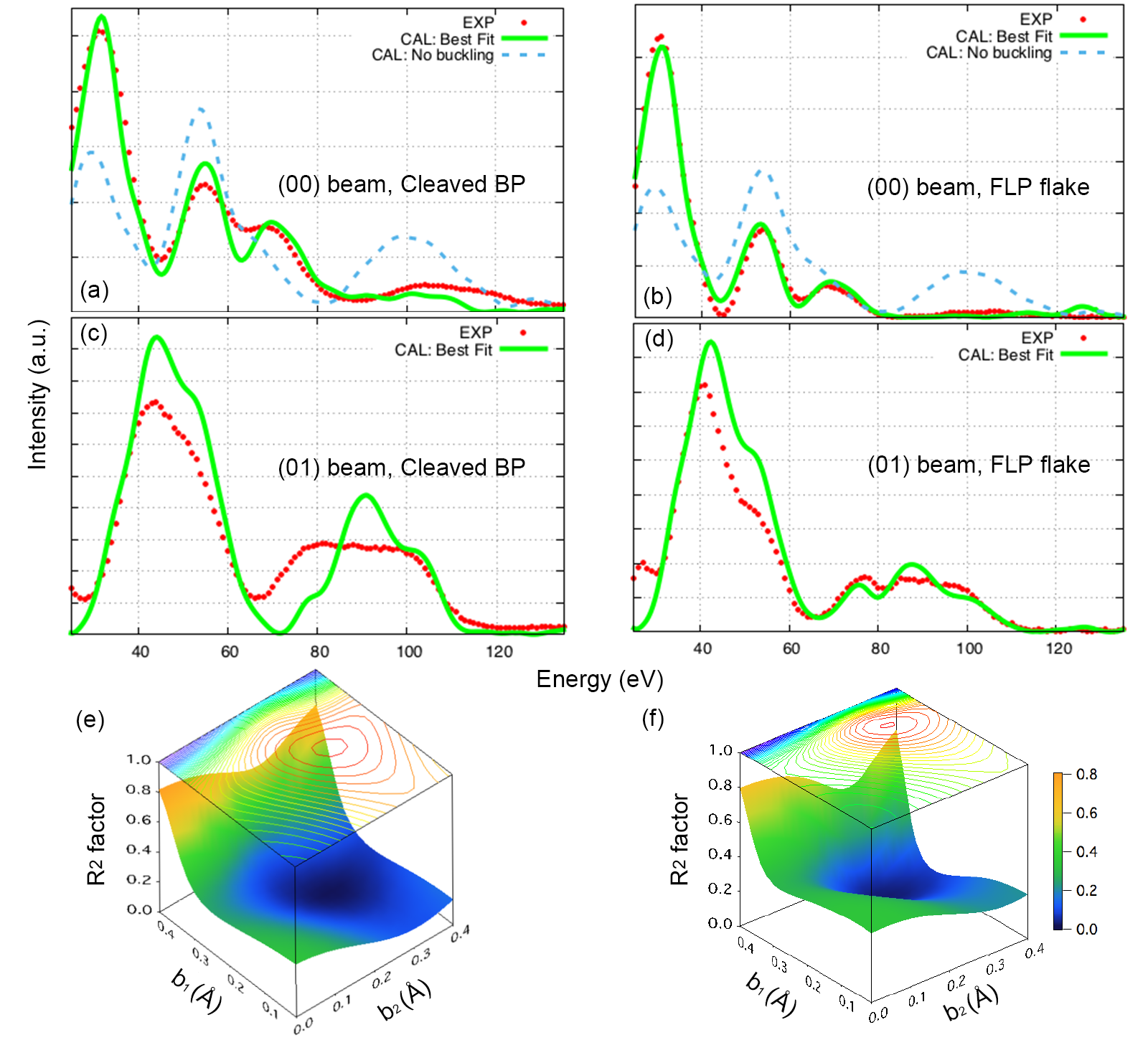}
\caption{(a) - (d): (00) and (01) low-electron energy diffraction beam \textit{IV} curves for cleaved BP crystal and exfoliated FLP flake, respectively. Green dotted curves are experimental and red line curves are calculated using optimized surface structural parameters. (e), (f):  Reliability \textit{$R_2$}-factor plotted vs. \textit{$b_1$} and \textit{$b_2$} for cleaved BP crystal and exfoliated FLP flake, respectively.}\label{fig:leediv} 
\end{figure*}

\begin{table*}[]
  \centering
\scriptsize
  \caption{Optimum parameter values for the surface structure of
  BP crystal and exfoliated BP flake}\label{tab:results}
\setlength{\tabcolsep}{1.0em} 
{\renewcommand{\arraystretch}{2.2}
\begin{tabular}{|l|c|c|c|c|c|c|c|}
  \hline 
  Model & T & $b_1$  (\AA) & $b_2$  (\AA) $$ &$z'$  (\AA) $$($\Delta{z}/z$)& $w'$  (\AA) $$($\Delta{w}/w$) \\
  
  \hline  \hline
  
  Cleaved BP & 300 K & \ \ \ \ \ \ \  0.225\ \ \ \ \ \ \ \ &\ \ \ \ \ \ \ \ $0.269$  \ \ \ \ \ \ \ \ &   2.287  (+5.3\%)&\ \ \ \ \ \ \ \ 2.825 (-8.0\%)   \ \ \ \ \ \ \ \ \\


 \hline 
 
 FLP & 573 K & \ \ \ \ \ \ \ \ 0.300 \ \ \ \ \ \ \ \ &\ \ \ \ \ \ \ \ $0.290$  \ \ \ \ \ \ \ \ & 2.381 (+9.9\%)     & \ \ \ \ \ \ \ \ 2.877 (-6.3\%)  \ \ \ \ \ \ \ \ \\
 

 \hline
  DFT \cite{stm1} & - & $0.02$ & -  & - & - \\
  
  \hline
  
     DFT \cite{stm2} & -  & $0.06$ & -  & - &  - \\
  
  \hline
  
  

\end{tabular}
}
\end{table*}

Multiple scattering theory and a muffin-tin potential model were implemented to calculate the LEED-\textit{IV} curves \cite{pendry,vanhove}. We used computer codes from Adams \textit{et al.} \cite{adam} which were developed from the programs of Pendry \cite{pendry} and Van Hove and Tong \cite{vanhove}. The utilization of the $R_2$ factor in Adams' codes allows the relative intensities of the diffraction beams to be preserved during the optimization, which enhances the reliability of the surface structure determination. The phase shifts (a quantity describing the atomic scattering property \cite{vanhove}) were calculated using the Barbieri/Van Hove phase shift calculation package \cite{phshift-link}. The muffin-tin radii for phosphorus atoms was set to \textit{r}$_{P}^{MT}$ = 2.099 a.u. and 12 phase shifts (\textit{L} = 11) were used for the LEED-\textit{IV} calculation. The in-plane lattice constant was \textit{$a_1$} = 3.313 \AA\ and  \textit{$a_2$} = 4.374 \AA, the thickness of the phosphorene layer was \textit{z} = 2.166 \AA$~$and the van der Waals distance between phosphorene layers was \textit{w} = 3.071 \AA$~$for the bulk, as indicated in Fig. \ref{fig:bulk} \cite{morita}. 

The mean-square atomic vibrational displacements $<$${u^2}$$>_{T}$ for the P atoms were calculated individually according to the relation between Debye temperature $\theta$$_{D}$ and $<$${u^2}$$>_{T}$ at the sample temperature of \textit{T}=300 K for bulk BP and  \textit{T}=573 K for the FLP flakes using the following equation \cite{pendry}:
\begin{equation}
<u^2>_{T}{ }= \frac{9\hbar^2}{m_ak_B\theta_{D}}\;\;\Big\{\frac{~T^2}{\theta_D^2}\int_{0}^{\frac{\theta_D}{T}}\frac{xdx}{e^x-1} + \frac{1}{4}\Big\}
\end{equation}
where $m_a$ is the atomic mass, $\hbar$ is the Planck's constant and $k_B$ is the Boltzmann constant.
The Debye temperature $\theta$$_{D}$ was set to 550 K \cite{debye}. The inner potential, \textit{V$_{0}$+ $i$V$_{im}$}, was set to be independent of energy. The real part \textit{V$_{0}$} was initially set to 8 eV and adjusted through \textit{$\Delta$V$_{0}$} during the fitting process while the imaginary part \textit{V$_{im}$} was fixed at 6 eV. 

The surface structural parameters \textit{$b_1$}, \textit{$b_2$}, \textit{w}' and \textit{z}' were varied in search for the optimum surface structure. Best-fit structural parameter values are listed in Table \ref{tab:results} and compared with previously reported results \cite{stm1,stm2}. The calculated LEED-\textit{IV} curves using optimized structural parameters match well with the experimental curves for both the BP crystal surface at 300 K and exfoliated FLP flake at 573 K, as shown in Fig. \ref{fig:leediv} (a)-(d). The minimized \textit{$R_{2}$} factors are 0.03 and 0.02, respectively. For comparison, the calculated \textit{IV} curves (blue dashed lines in Fig. \ref{fig:leediv} (a)-(b)) using a flat, un-buckled surface, are distinctively different from our experimental results. For the freshly cleaved BP crystal surface, our results show that the top layer surface buckling \textit{$b_1$} is 0.22 \AA\ and the second phosphorus atomic layer buckling \textit{$b_2$} is 0.27 \AA. The thickness of the top phosphorene layer \textit{z}' is expanded by 5.3\% from its bulk value of 2.166 \AA. The van der Waals gap between the top and second phosphorene layer \textit{w}' is contracted by 8\% from its bulk value of 3.071 \AA. For the mechanically exfoliated flake of FLP at 573 K, the top and second layer buckling are 0.30 \AA\ and 0.29 \AA, respectively. The surface bucklings are slightly larger at 573 K than the BP crystal surface at 300 K. We attribute this increase of surface buckling to thermal surface expansion at elevated temperature. For the same reason, the top phosphorene layer \textit{z}' and the top van der Waals gap \textit{w}' are also slightly increased at 573 K compared to 300 K. \textit{z}' shows an expansion of 9.9\% and \textit{w}' a contraction of 6.3\%, with respect to their corresponding bulk values. Figure \ref{fig:leediv} (e)-(f) show plots of the reliability \textit{$R_2$} factor as a function of the surface buckling \textit{$b_1$} and the second atomic layer buckling \textit{$b_2$} for both of the investigated samples. Well-defined minima were observed for both cases. Along with the good agreement of experimental and calculated \textit{IV} curves, both results give us strong confidence in our findings.

Our most striking result is that the BP surface buckling \textit{$b_1$} is one order of magnitude larger than the previously proposed theoretical values \cite{stm1,stm2}, for both BP and FLP samples investigated. Note that the buckling extends to second atomic layer. Similar significant surface buckling has also been predicted for other group V thin film such as Bi and other similar elemental 2D materials such as silicene, germanene, by various first-pricinples studies. Specifically, Cahangirov et al. predicted that the buckling height for silicene to be 0.44 \AA\ and 0.64 \AA\ \cite{silicene2}; our co-author, Sadowski et al., proposed the buckling of Bi thin film to be 0.5 \AA\ \cite{bi1, bi2}. In order to further support our surface structural results, first-principles calculations were utilized to investigate the origin of this surface reconstruction as well as its influence on the electronic properties of BP and FLP. 


First-principles studies of the surface structure were carried out based on the framework of density functional theory (DFT) with projector augmented plane-wave (PAW) potential \cite{PAW} as implemented in the Vienna ab initio simulation package (VASP) \cite{VASP1,VASP2,VASP3}. The plane-wave functions expanded with an energy cutoff of 400 eV were employed throughout calculations. The exchange-correlation energy was described by generalized gradient approximation (GGA) in Perdew, Burke, and Ernzerhof (PBE) form \cite{PBE}. The \textit{k} points in two-dimensional Brillouin zone (BZ) of the 1$\times$1 unit cell of monolayer BP containing 4 phosphorous atoms were sampled on a 16$\times$12 mesh. The van der Waals (vdW) interactions were also incorporated within the Tkatchenko-Scheffler method \cite{vdW}. In addition, we employed the Heyd, Scuseria, and Ernzerhof (HSE06) hybrid functional \cite{HSE1,HSE2} for the band structure calculations.

The structure of monolayer phosphorene and the top phosphorene layer of bulk BP (a six-layer supercell) were calculated and compared. Only very small structural differences were observed, $<$ 0.001 \AA, between the atomic positions and and bond lengths of the monolayer phosphene and that of the top layer of bulk. This is expected for layered materials with weak van der Waals bonding in between adjacent layers. In order to simplify our calculations, we focus on single layer phosphorene. The thickness of the vacuum layer in each slab structure is more than 15 \AA.


First, the defect-free monolayer phosphorene with different sizes were investigated. The lattice structure was optimized until the atomic force, both Hellmann-Feynman and vdW terms included, on each relaxed atom was less than 1 meV/\AA. In an up to 8$\times$4 supercell, no buckling was found within the accuracy of the calculation. This result is reasonable since both BP bulk and monolayer structures have the insulating electronic structures with band gaps, and exposed surfaces do not bring about the electronic mismatch or additional dangling bonds. Surface reconstruction is thus not necessary in such a stable structure.

\begin{figure}
\centering     
\includegraphics[width=80mm]{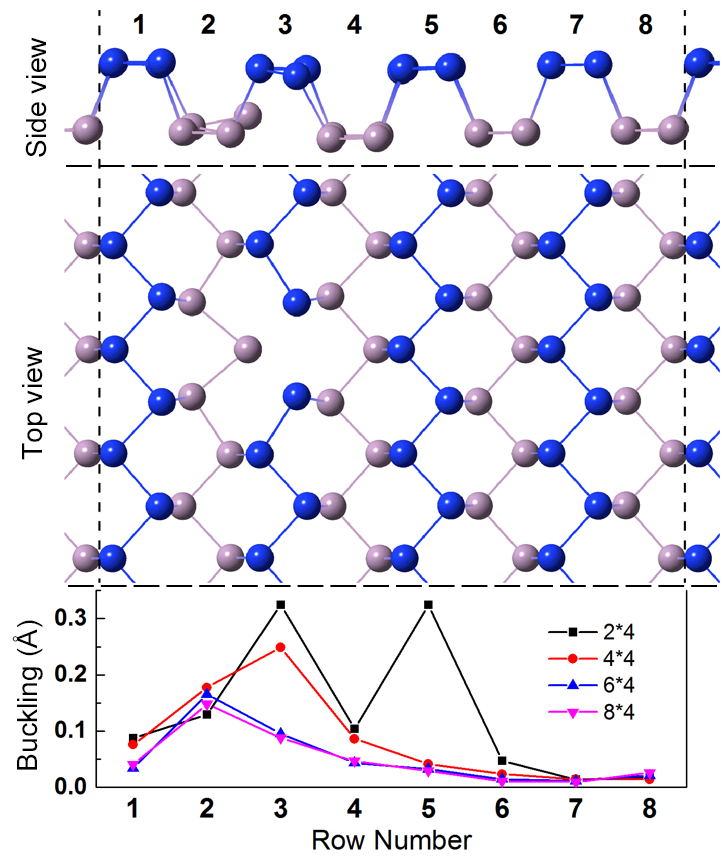}
\caption{Phosphorene atomic structure with defect introduced. Upper pannels: Side and the top view of a n$\times$4 (n=2, 4, 6, 8) supercell of the monolayer phosphorene with a point defect introduced at row 3.  Blue and grey color of balls distinguish the top and second P atomic layers. Lower panel: Average magnitude of buckling in each row for various n$\times$4 supercells. 
}\label{fig:dft-defect}
\end{figure}

However, if an impurity, such as vacancy defect \cite{stm2} or doping \cite{bp3, ptype-1} is induced on the surface, the situation changes completely. In fact, Liang \textit{et al.} have recently observed vacancy defects on their freshly cleaved surfaces of BP crystals \cite{stm2} using STM. Here, we introduced a single point defect into the monolayer phosphorene by removing one atom. Several supercells were calculated with their sizes ranging from 2$\times$4 to 8$\times$4. After the structure optimization, deviations of the atoms along out-of-plane direction were observed in all of these structures. As shown in the top view (middle panel) of Fig. \ref{fig:dft-defect}, each supercell has 8 zig-zag rows, and the defect is located on the upper layer of row 3. The magnitude of buckling in each row is summarized in the bottom panel of Fig. \ref{fig:dft-defect}. by calculating the standard deviations of the phosphorus atoms' \textit{z} components for each entire row. It is seen that the buckling is maximized in rows around the defect, and the maximum buckling ranges from 0.15 \AA\ to 0.33 \AA\ through all the supercell sizes under investigation. These calculations agree well with our experimental values of 0.22 \AA\ to 0.30 \AA. Although the buckling magnitude decays rapidly along the armchair direction, away from the row, on which the defect is located, no significant decay in the buckling magnitude was found in the zig-zag direction. Based on these results, it is concluded that the buckling is significantly enhanced near the point defect. It is anisotropic and long-range along the zig-zag direction while it is short-range along the armchair direction. The defect-induced buckling cannot be maintained in the armchair direction. This interesting insight agrees well with the recent experimental observation of an anisotropy in the surface density-of-state (DOS) on the BP surface by STM \cite{stm2}. 

\begin{figure}
\centering     
\includegraphics[width=80mm]{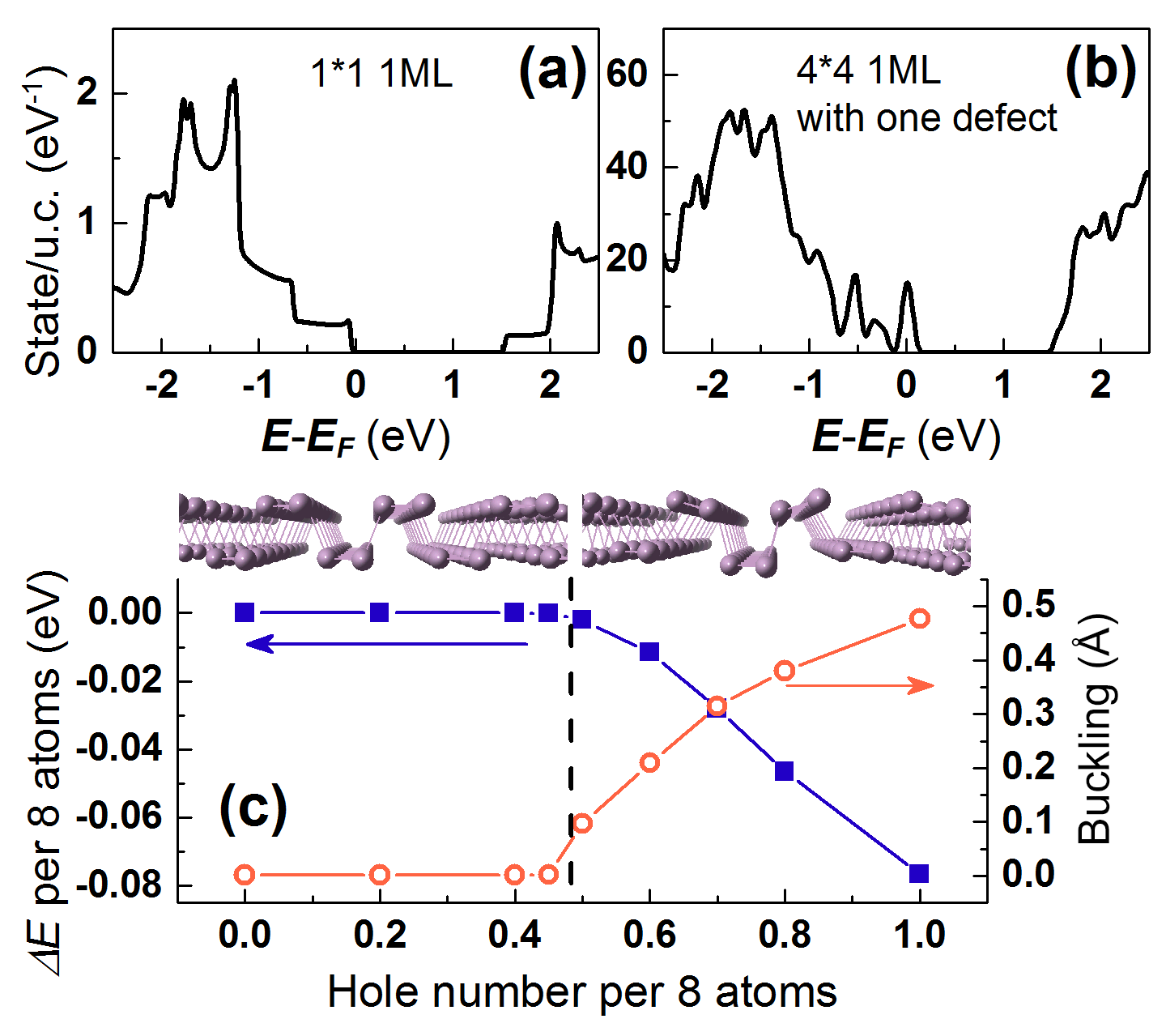}
\caption{Buckling and hole-doping induced by defects. The DOS for (a) the ideal monolayer and (b) the 4$\times$4 defect-included supercell BP. The Fermi level is set to zero. (c) Energy difference (blue solid squares) between the buckled and non-buckled configurations and the magnitude of buckling (red open circles) as the increasing hole-doping number.
}\label{fig:dft-doping}
\end{figure}

Intuitively one would expect that such long-range  buckling would be reflected in the band structures as well. Thus the electronic structure of the 4$\times$4 supercell with a single vacancy was investigated and compared with that of a clean monolayer. According to the density-of-state results shown in Fig. \ref{fig:dft-doping} (a)-(b), the clean monolayer BP is insulating with a band gap of 1.5 eV, while an impurity state is present in the defect containing supercell across the Fermi level close to the top of valence states. A similar state was also observed by Zhang et al. \cite{stm1} in their STM \textit{dI/dV} measurement. This indicates the existence of the defect-induced hole-doping electronic structure in these defect structure, i.e. each phosphorus vacancy generates three dangling bonds that need to be saturated by more electrons. This suggests that the distortion of the lattice, such as buckling, appears in order to eliminate this instability of the electronic structure.  

To better understand the relation between hole-doping and the surface structure of BP, the hole doped 2$\times$1 clean supercell structures with a tunable total electron number was optimized. As shown in Fig. \ref{fig:dft-doping} (c), buckling appears when the hole number exceeds 0.5 per 8 phosphor atoms. The magnitude of the buckling as well as the energy difference between the buckled and ideal structures increases rapidly with the rise of the hole number. In particular, the buckling reached 0.2 \AA\ when the hole number is 0.6 per 8 atoms. Our first principle calculations thus show that the presence of defect induces hole doping on the clean BP surface, which in turn leads to lattice distortion and the surface buckling. It was confirmed experimentally that both undoped bulk BP \cite{ptype-1} and FLP \cite{bp3} are p-type semiconductors, but the origin of intrinsic p-type doping is unclear so far. Recently, Osada proposed that the edge state of finite bi-layer phosphorene might be the origin of the intrinsic hole-doping around the edge \cite{ptype-2}. Our DFT calculations, together with the experimental observation of BP surface reconstruction, strongly indicate that surface defect induced buckling is a likely explanation for the intrinsic hole-doping for both bulk BP and FLP. 

To summarize, we determined the surface atomic structure of black phosphorus using dynamical selected area $\mu$LEED-\textit{IV} in a high spatial resolution LEEM system. Two samples were studied at two different temperatures; freshly cleaved bulk BP at room temperature and mechanically exfoliated few-layers phosphorene at 573 K. The results show that for both samples, the surface buckling is one order of magnitude larger than previously proposed. The surface buckling on black phosphorus was found to be profound, and is close to the buckling of other similar elemental 2D materials such as silicene and germanene \cite{silicene2} and other group V thin film materials such as Bi \cite{bi2}. Specifically, the top surface buckling \textit{$b_1$} for the cleaved BP crystal surface at 300 K is 0.22 \AA\ and for exfoliated FLP of about 10 nm thickness at 573 K is 0.30 \AA. The slight increase in the buckling magnitude from 300 K to 573 K is most likely due to the thermal expansion. Similar phenomenon has been observed on other 2D materials, such as MoS$_2$ \cite{daileed}. Furthermore, due to the high sub-surface sensitivity of $\mu$LEED-\textit{IV}, we found a similar buckling for the second phosphorus atomic layer. We denoted this buckling as \textit{$b_2$}=0.27 \AA\ and 0.29 \AA\ for T=300 K and 573 K, respectively. Besides the surface buckling, our results show that the BP surfaces have an expansion-contraction surface relaxation. The thickness of the top phosphorene layer \textit{z}' is expanded by 5.3\% in the cleaved BP crystal at 300 K and expanded by 9.9\% in the exfoliated FLP at 573 K. The van der Waals gap between the first and second phosphorene layer \textit{w}' is contracted by 8\% in cleaved BP crystal at 300 K and ccontracted by 6.3\% in exfoliated FLP at 573 K.  

We further confirmed our surface structural results using first-principles calculations and identified a vacancy defect driven mechanism as the cause of the surface buckling. The surface vacancy defect also introduces an impurity state in the band gap, which suggests that the vacancy defect induced surface buckling is most likely the reason why phosphorene materials are intrinisically p-type.

We have shown that $\mu$LEED-\textit{IV} via LEEM, with its high surface sensitivity and selected area investigation ability, is a powerful technique to study atomic lattice of 2D materials. We believe that our results provide a clear picture of the fundamental surface properties of BP, including the detailed surface structure, the origin of its surface deformation, the electronic properties of phosphorene materials, and the origin of its intrinsic hole-doping property. We expect the significant surface buckling to potentially be even more pronounced in the monolayer phosphorene form, and having a significant impact on the electronic properties of monolayer phosphorene based devices.

\begin{acknowledgement}
The experimental work presented here was carried out at the Center for Functional Nanomaterials, which is a U.S. DOE Office of Science Facility, at Brookhaven National Laboratory under Contract No. DE-SC0012704. The work of Z. D. and K. P. were supported by NSF DMR 1006863. First principles analysis by J.X.\ Y.\ and J. Z. was supported by the U.S. Department of Energy (DOE), Office of Science, Basic Energy Sciences (BES) under Award No. DE-SC0016424.  In addition, the work of R.M.\ O. and W. J. were financially supported by the U.S. Department of Energy under Contract No. DE-FG 02-04-ER-46157. The work of J. H. and Y. D. K. were financially supported by NSF MRSEC program through Columbia in the Center for Precision Assembly of Superstratic and Superatomic Solids (NSF DMR-1420634). 
\end{acknowledgement}


\bibliography{bp}








\end{document}